\documentclass{eds2019}
\usepackage{cite}

\def\Journal#1#2#3#4{{#1} {\bf #2}, #3 (#4)}

\def\NPA{{\em Nucl. Phys.} A}
\def\PLB{{\em Phys. Lett.}  B}
\def\PRL{\em Phys. Rev. Lett.}
\def\PRD{{\em Phys. Rev.} D}
\def\PRC{{\em Phys. Rev.} C}

\def\IJMPE{{\em Int. J. Mod. Phys.} E}
\def\JHEP{\em JHEP}
\def\EPJC{{\em Eur. Phys. J.} C}
\def\PoSHP{\em PoS HardProbes}

\def\be{\begin{equation}}
\def\ee{\end{equation}}
\def\bea{\begin{eqnarray}}
\def\eea{\end{eqnarray}}

\newcommand{\Photo}{\includegraphics[height=35mm]{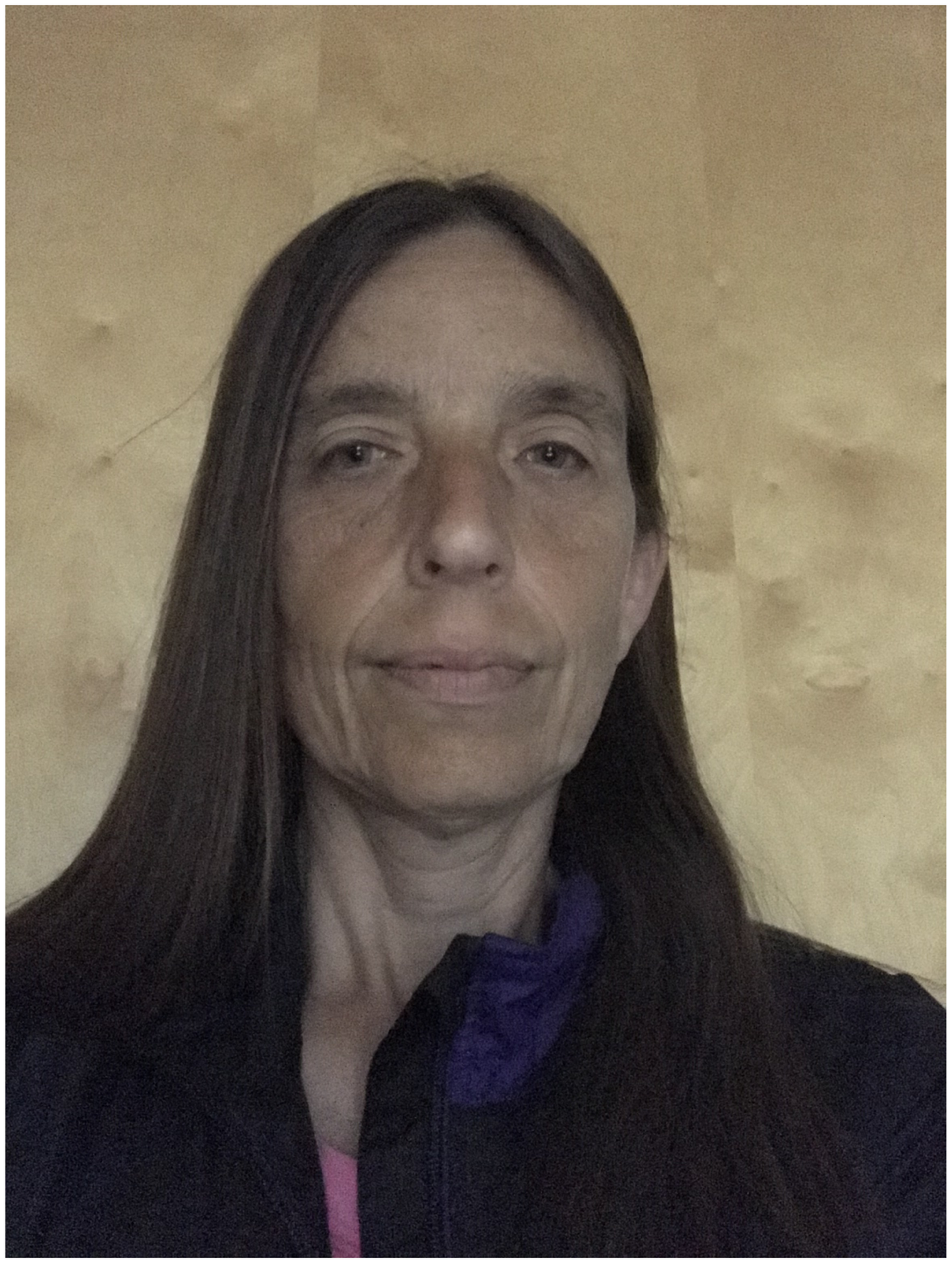}}

\begin{document}
\vspace*{4cm}
\title{REVIEW OF PREDICTIONS OF HARD PROBES IN $p+$Pb COLLISIONS
  AT $\sqrt{s_{NN}} = 5.02$ AND 8.16~TeV AND COMPARISON WITH DATA
  }

\author{R. Vogt}

\address{Lawrence Livermore National Laboratory, 
Livermore, CA
94551, USA \\ 
Physics Department, University of California, Davis, CA 95616,
USA}

\maketitle\abstracts{Predictions have been compiled for the $p+$Pb LHC runs,
  focusing on production of hard probes in cold nuclear matter.  These
  predictions were first made for the $\sqrt{s_{_{NN}}} = 5.02$ TeV $p+$Pb run
  and were later compared to the available data.  A similar set of predictions
  were published for the 8.16~TeV $p+$Pb run.  A selection of the predictions
  are reviewed here.
}

\section{Introduction}

This proceedings paper covers some of predictions for the production of hard
probes in minimum bias $p+$Pb collisions in the $\sqrt{s_{_{NN}}} = 5.02$ and
8.16~TeV runs in 2012 and 2016.  The predictions at 5.02~TeV were presented in
Ref.~\cite{Pred1} with a follow up comparison to the data published so
far in Ref.~\cite{Pred2}.  The predictions for 8.16~TeV were compiled in
Ref.~\cite{Pred3}.  The focus was on hard probes because they are high
mass or
high energy probes and therefore calculable in perturbative QCD.  Due to their
higher energy scales, they are produced early in the history of the collision,
thus carrying information about the state of the system when they were produced.
This is especially true of probes such as hard photons, Drell-Yan dileptons and
massive gauge bosons which are unaffected by the strong interaction and thus
travel through the medium without interaction.  They are thus especially
important for differentiating the parton distributions in a nucleus from those
in a proton.

The two proton-lead runs at the LHC have provided access to a system that is
intermediate to the ``vacuum'' of proton-proton collisions and the hot dense
quark-gluon plasma produced in heavy-ion collisions such as Pb+Pb.  There have
been important comparisons between $p+$Pb and $p+p$ collisions to determine
the level of ``cold nuclear matter'' effects, the modifications of hard probes
due to the nuclear medium without a quark-gluon plasma, while comparisons
between $p+$Pb and Pb+Pb collisions differentiates between cold and ``hot''
nuclear matter.  While the calculations discussed here were made for minimum
bias collisions, with a relatively low multiplicity, it has been noted that the
highest multiplicity $p+p$ and $p+$Pb collisions share some characteristics
with heavy-ion events.  For a more detailed discussion of this and additional
references, see Ref.~\cite{Pred3}.

Due to the lack of space, there can be only a minimal discussion of the
predictions.  The fewest required references are included here, in particular
any new data since the publication of the compilations in
Refs.~\cite{Pred1,Pred2,Pred3}.
Please see the compilations themselves for full details, along with references
to the original work.

The focus here is on new results.  Therefore, under quarkonium and heavy
flavor, new data on $\Upsilon$ \cite{ALICE_ups8,LHCb_ups8} from ALICE and
LHCb and $B$ mesons from LHCb
\cite{LHCb_NPjpsi,LHCb_Bplus} modifications at 8.16~TeV are compared to
predictions. New modifications of the nuclear parton densities based on the
dijet and gauge boson data, the EPPS16 \cite{epps16} set, is discussed,
followed by a discussion of these data at 5.02~TeV.  Finally, top quark
production, measured in collisions involving nuclei for the first time, is also
discussed.

\section{$\Upsilon$ and $B$ meson modifications}

Quarkonium and open heavy flavor production were presented together in the
compilations.  In Refs.~\cite{Pred1,Pred2,Pred3}, $J/\psi$ production was
calculated both in approaches employing collinear factorization and saturation
approaches.  Since $\Upsilon$ results are shown here, only results assuming
collinear factorization are shown because the $\Upsilon$ mass scale is too high
for saturation effects to be relevant.

\begin{figure}[htbp]
\begin{center}
\includegraphics[width=0.45\textwidth]{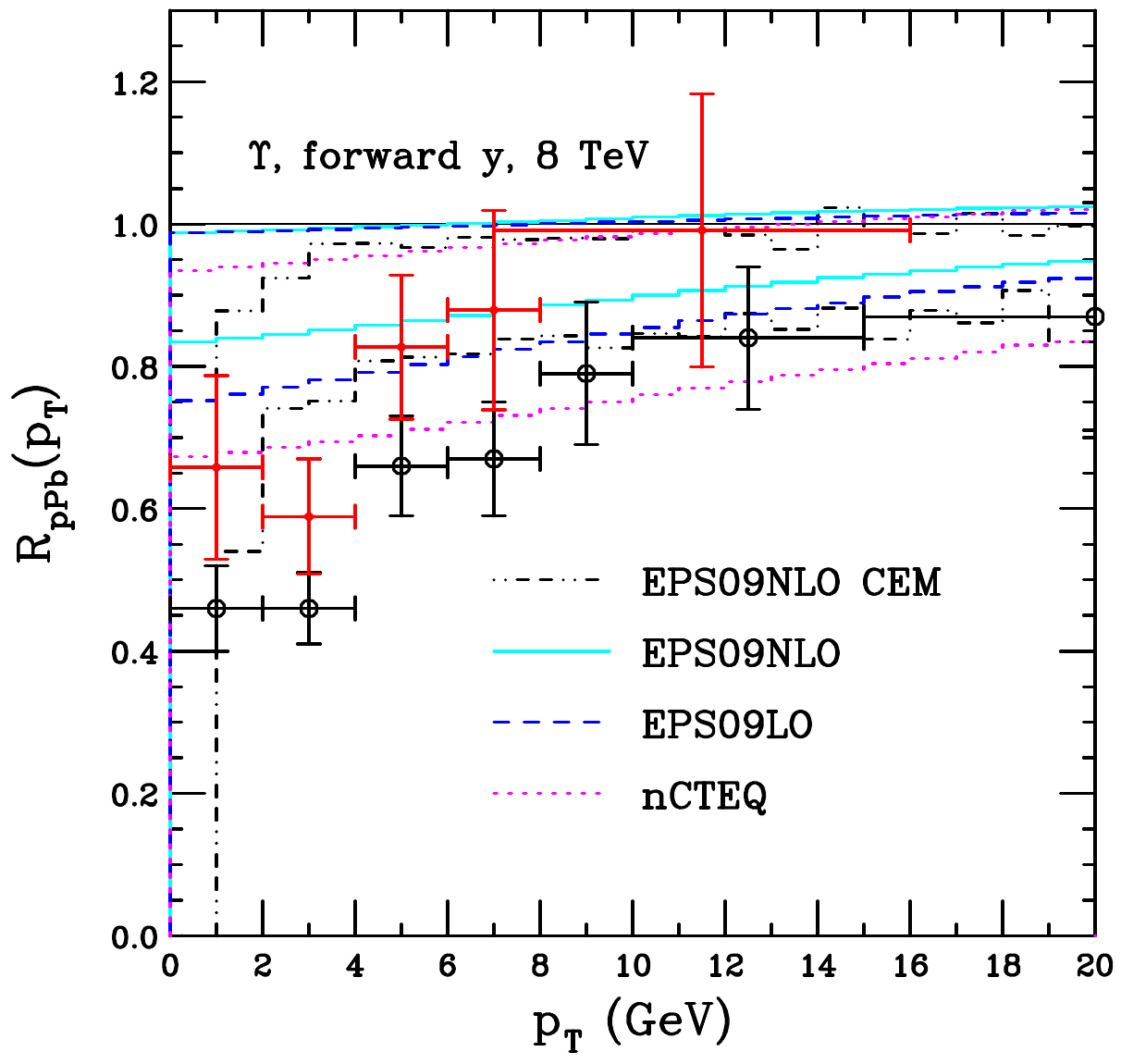}
\includegraphics[width=0.45\textwidth]{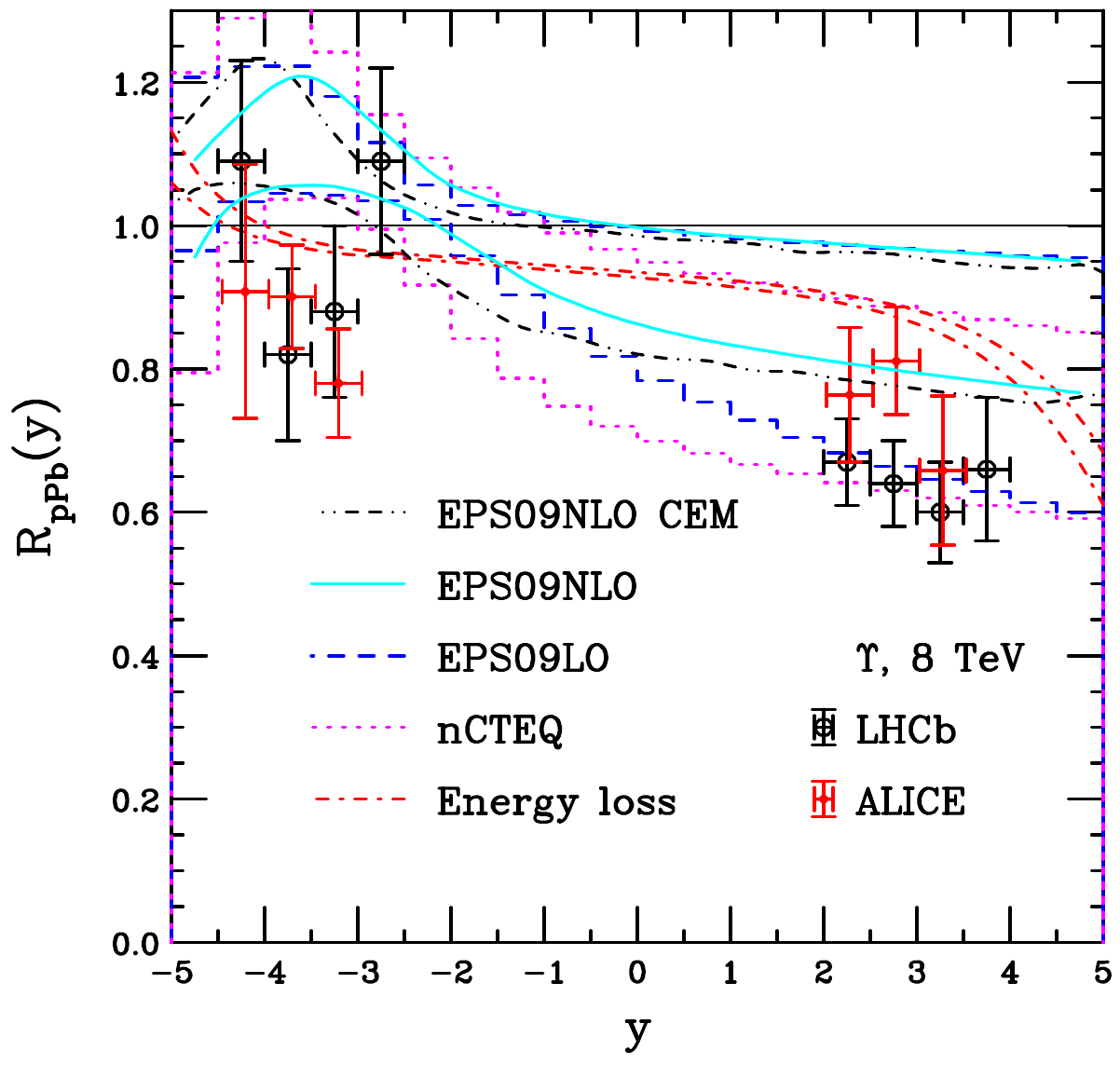}
\end{center}
\caption[]{
  The nuclear suppression factor $R_{\rm p{\rm Pb}}$ for $\Upsilon$ at
  $\sqrt{s} = 8.16$~TeV as a function of $p_T$ at forward rapidity (left)
  and as a function of rapidity (right).  The EPS09 NLO results in the CEM
  (dot-dot-dash-dashed black) are shown with Lansberg and Shao's data-driven
  fits with EPS09 NLO (solid cyan), EPS09 LO (dashed blue) and nCTEQ
  (dotted magenta). On the right hand side, an energy loss calculation 
  by Arleo (dot-dashed red curve) is also shown.
  The LHCb \protect\cite{LHCb_ups8} and 
  ALICE \protect\cite{ALICE_ups8} data are shown by the black and red points
  respectively.  Modified from Ref.~\protect\cite{Pred3}.
}
\label{fig:Upsilon}
\end{figure}

The calculations shown in Fig.~\ref{fig:Upsilon} are predominantly from two
sources, both of which focus on nuclear PDFs and are shown as functions of
transverse momentum, $p_T$, and rapidity, $y$.  An additional calculation, based
on energy loss, without any nPDF modification, is only shown as a function of
rapidity.  The calculations labeled EPS09NLO CEM are made in the color
evaporation model, CEM,
at next-to-leading order, NLO, with the EPS09 nPDF set using
$b$ quark masses and scales from a fit to $b \overline b$ cross section data
\cite{Vogt:2015uba}.
The calculations labeled EPS09NLO, EPS09LO and nCTEQ \cite{Kovarik:2015cma}
were made by Lansberg and Shao \cite{Lansberg:2016deg}
employing a data-driven approach featuring the $gg$ channel only with
parameterized amplitude and coefficients that can be fit to $p+p$ data.  Note
that the EPS09NLO calculations are similar in the two approaches but not
identical, the CEM calculation is a complete NLO calculation, with all
production channels include and no a prior assumption that the factorization
and renormalization scales need to be equal to the mass.  The energy loss
calculation by Arleo, shown as a function of rapidity, also assuming a
parameterized fit to the $p+p$ cross section.  The energy loss is implemented by
a shift of the rapidity in $p+$Pb collisions relative to $p+p$ collisions.

The trends of all the nPDF calculations are similar.  At forward rapidity,
there is a depletion of the nuclear modification factor, $R_{p {\rm Pb}}$, at
low $p_T$.  (Here, and in the rest of this paper, $R_{p {\rm Pb}}$ is effectively
calculated as the cross section per nucleon in $p+$Pb collisions relative to
the $p+p$ cross section in the same kinematic range.)  The depletion goes away
at higher $p_T$ due to the nPDF scale evolution. As a function of rapidity,
the calculations show antishadowing at negative rapidity, larger momentum
fractions, $x$, in the lead nucleus, and a depletion at forward rapidity, where
the $x$ in the lead nucleus is low.  There are large uncertainty bands
reflecting the error sets of the nPDF global analyses.  Because the energy loss
calculation does not assume any difference between the gluon PDFs in the proton
and the lead nucleus, the only uncertainty is due to that on the energy loss
parameter, resulting in a narrow range of energy loss predictions.  All of the
calculations are within the uncertainties of the data from ALICE
\cite{ALICE_ups8} and LHCb \cite{LHCb_ups8}, shown in the black and red points
respectively.

\begin{figure}[htbp]
\begin{center}
\includegraphics[width=0.45\textwidth]{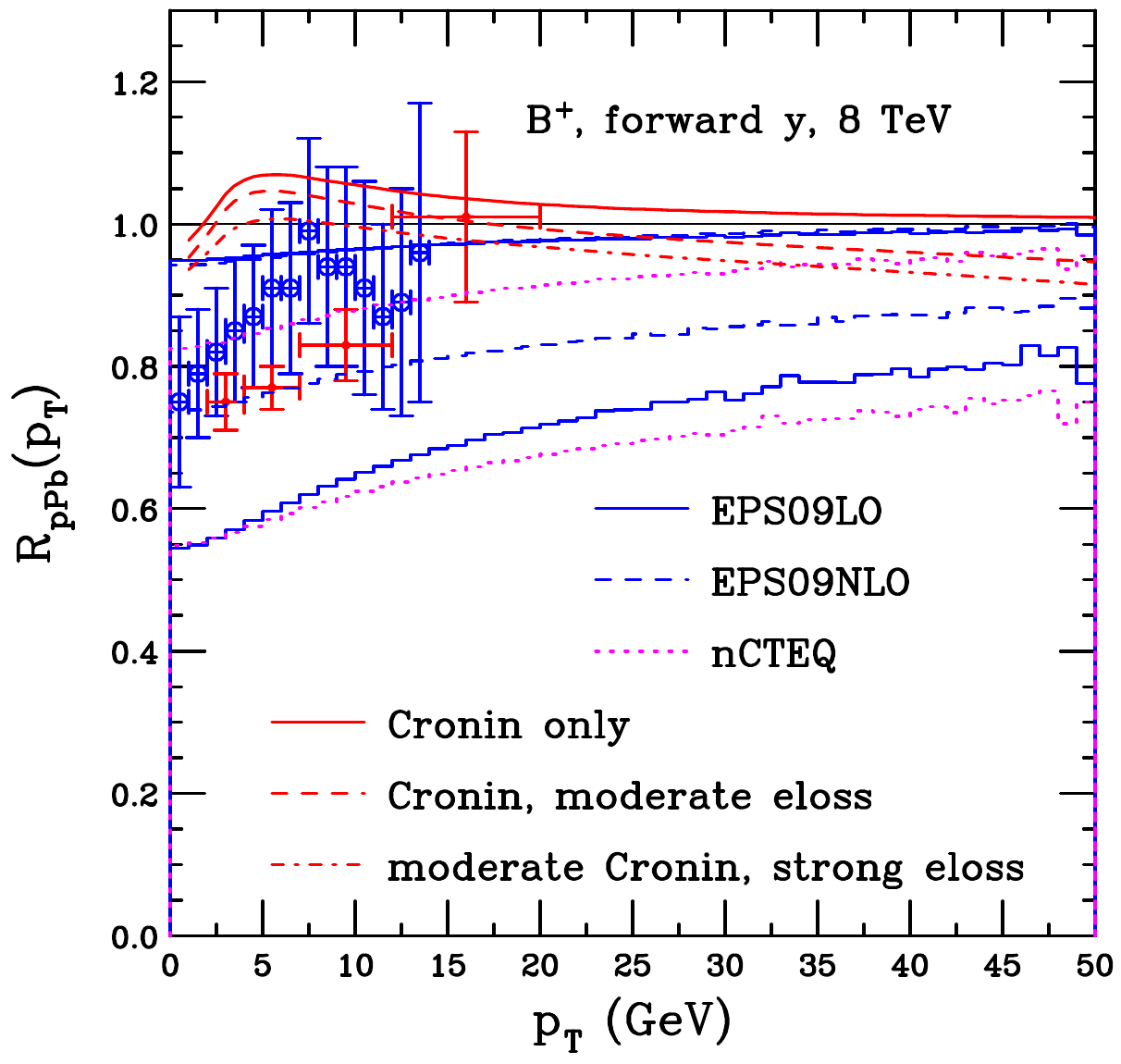}
\includegraphics[width=0.45\textwidth]{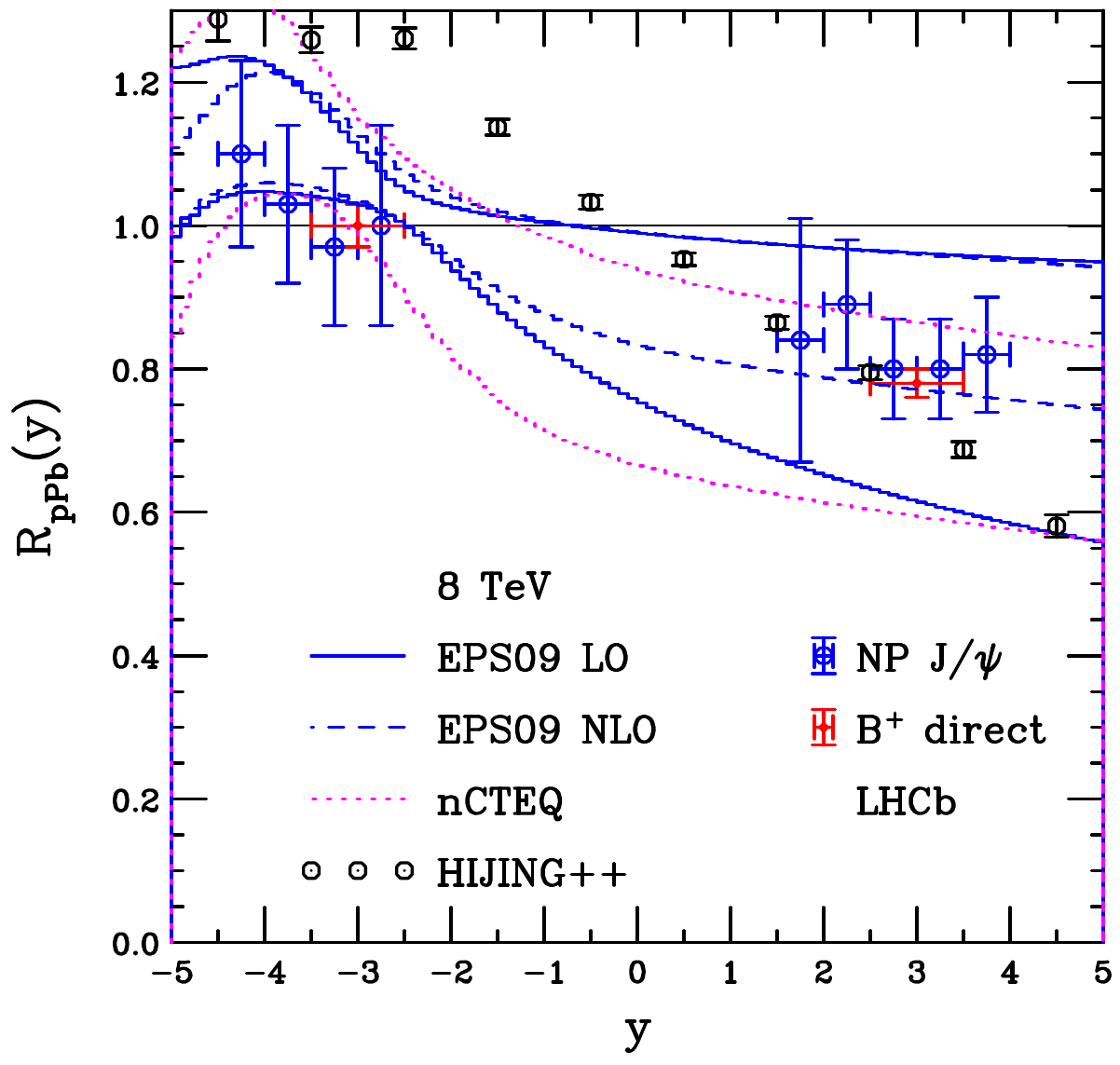}
\caption[]{
  The calculated $R_{p{\rm Pb}}$ for the LHC non-prompt $J/\psi$
  \protect\cite{LHCb_NPjpsi} 
  and $B^+$ \protect\cite{LHCb_Bplus} 
  data are compared with with EPS09 LO (blue),
  EPS09 NLO (cyan) and nCTEQ (red) as a function of $p_T$ at forward rapidity
  (left) and as a function of rapidity (right).  Also shown as a function of
  $p_T$ is a prediction with $k_T$ broadening (Cronin) and energy loss while a
  prediction of $\mathtt{HIJING++}$ is shown as a function of rapidity.
  Modified from Ref.~\protect\cite{Pred3}.
}
\label{fig:Bplus}
\end{center}
\end{figure}

The same data driven parameterization, albeit with different parameter values,
are also shown in Fig.~\ref{fig:Bplus} for $B^+$ production in 8~TeV collisions.
The trends are the same as those seen as a function of $p_T$ and $y$ for
$\Upsilon$ production.

Two additional predictions are shown for $B$ mesons.  The calculations of
Vitev and collaborators, see Refs.~\cite{Pred2,Pred3} for details,
labeled as `Cronin' and `eloss' on $R_{p{\rm Pb}}(p_T)$,
also include dynamical shadowing, different from the nPDF parameterizations.
The Cronin effect results in $k_T$ broadening
and thus an enhancement at low $p_T$, rather than a depletion due to the nPDF
only calculations by Lansberg and Shao.  The inclusion of energy loss in the
calculations weakens the low $p_T$ enhancement.  The calculations shown as a
function of rapidity also include a prediction from $\mathtt{HIJING++}$ a
revised version of the general purpose $\mathtt{HIJING}$ simulation of heavy-ion
collisions, shown as points in the figure.

LHCb data on non-prompt $J/\psi$, those from $B$ meson decays, as well as data
where the $B^+$ is reconstructed \cite{LHCb_NPjpsi,LHCb_Bplus}.  The data are
most consistent with the nPDF calculations.  The calculations assuming Cronin
and energy loss are more inconsistent with the data since they do not show any
enhancement at low $p_T$.  $\mathtt{HIJING++}$ is too high at negative rapidity
but is consistent with the calculations at forward rapidity.

These results, with the large nPDF uncertainties, highlight the need for updated
nPDF sets, especially in further constraining the gluon nPDFs.

\section{Modifications of the Parton Densities in Nuclei}

One physics outcome from the 5~TeV $p+$Pb run was the new set of nuclear
parton distribution functions (nPDFs) by Eskola and collaborators, EPPS16
\cite{epps16}.
This set is the first to include the LHC data, specifically
that of $W^\pm$ and $Z^0$ production from
CMS \cite{WpWm_5TeV,Z0_5TeV}
and ATLAS \cite{Aad:2015gta} as well as the dijet
data from CMS \cite{Chatrchyan:2014hqa}.  
These data could be included in the global analyses because they were all
forward-backward asymmetry data and do not rely on a $p+p$ baseline at the same
energy.  They also added, for the first time for the Eskola {\it et al.}
sets, neutrino deep-inelastic scattering data.

Incorporating the LHC and neutrino data into the analysis allowed more detailed
flavor separation for the quark sets.  In particular, the LHC data allowed
them to increase the fit range in
momentum fraction, $x$, and factorization scale, $Q^2$,
to regions heretofore
unavailable in lepton-nucleus collisions.
Unfortunately, even with the dijet data from CMS, the gluon
distribution in the nucleus, particularly at low $x$ and moderate
$Q^2$, is still not well constrained.

These sets were not yet available at the time most of the predictions for
Ref.~\cite{Pred3}
were collected except for the top quark predictions, shown
later.  However, it is worth noting that the central EPPS16 set
gives results
quite similar to those calculated with EPS09 NLO \cite{eps09}.
The largest change, for
gluon-dominated processes, is the increase in the uncertainty band due to the
increased number of parameters required for flavor separation and the relaxing
of some previous constraints.  See Ref.~\cite{epps16}
for details and comparison to the 5.02 TeV $p+$Pb data included in the global
analysis.

One might expect further global analyses of the nuclear parton densities after
more data, especially from the 8.16 TeV run become available.
At a given $p_T$, the $x$ value probed in
a hard scattering process is a factor of 0.62 smaller at 8.16~TeV than 5.02~TeV.
In addition, the higher energy allows a somewhat broader reach in rapidity so
that some processes, such as $Z^0$ production at LHCb, see the discussion in
Ref.~\cite{Pred2}, measured near the edge of
phase space, can expect higher statistics and perhaps high enough significance
to be included in future global fits.  Similarly, the $p_T$ reach of most
processes is increased.

\section{Dijets}

The first dijet data from CMS were binned in dijet pseudorapidity,
$\eta_{\rm dijet}$, defined as
half the sum of the pseudorapidities of the leading and subleading jets, the
two hardest (highest $p_T$) jets in the event.  It was seen that the dependence
of the normalized dijet distribution tracks the $x$ dependence of the nPDFs.
Because of the high $p_T$ scales, $p_T > 120$~GeV for the leading jet and
$p_T > 30$~GeV for the subleading jet, at $\sqrt{s_{_{NN}}} = 5.02$~TeV, with
$\eta_{\rm dijet} < 0$ in the laboratory frame, the EMC region, $x > 0.3$, is
probed while, at forward $\eta_{\rm dijet}$, the antishadowing region,
$0.03 < x < 0.3$, is studied.

In Ref.~\cite{Pred2}, the CMS 
$p_T$ integrated dijet results \cite{Chatrchyan:2014hqa}
were compared to two calculations: the CT10 proton PDFs \cite{Lai:2010vv}
alone and CT10 with
EPS09 NLO.  When EPS09 NLO was included, the ratio data/EPS09 was within the
uncertainty of the nPDF sets while the same ratio with CT10 alone shows
significant discrepancies \cite{Eskola:2013aym}.

CMS has recently published a more thorough analysis of these
data \cite{CMS_dijets_new}.  The ratio
$p$Pb/$pp$ was given in five
bins for the average $p_T$ of the two jets in the dijet:
$55 < p_T^{\rm ave} < 75$~GeV; $75 < p_T^{\rm ave} < 95$~GeV;
$95 < p_T^{\rm ave} < 115$~GeV; $115 < p_T^{\rm ave} < 150$~GeV; and
$p_T^{\rm ave} > 150$~GeV and compared to different nPDF sets available before
the LHC $p+$Pb runs.  (Note that these ratios were not directly available for
the EPS09 global analysis, only the forward-to-backward pseudorapidity ratio
could be used because the $p+p$ data at 5~TeV were taken only after the 2012
$p+$Pb run at the same energy.)
While none of these data sets probe the shadowing region
deeply, because gluons dominate jet production, these data provide the
greatest insight into the nuclear gluon
distribution in the range $0.003 < x < 1$ at the highest $p_T$ scales so far
available.  This is indeed a great advance because nuclear deep-inelastic
scattering can probe gluon distributions only indirectly through their scale
evolution and other, lower mass probes of the nuclear gluon density, such as
quarkonium, suffer from uncertainties regarding the production mechanism and
the relative importance of other cold nuclear matter effects.

While EPS09 still gives a better description of the CMS data than either DSSZ
\cite{DSSZ} or nCTEQ,
EPPS16, which benefited from a global analysis including the
$p_T$-integrated dijet data \cite{Chatrchyan:2014hqa}, gives a superior result
to EPS09 compared to the results in different $p_T$ intervals.

\section{Gauge Bosons}

Massive gauge boson production also provides new and important insight into
the charged parton distributions in nuclei, including any differences in the
up and down sea quarks, especially for $W^\pm$ production and the corresponding
charge asymmetry.  Heretofore, only Drell-Yan data in fixed-target
experiments were available to probe this difference.  While the scale probed
is somewhat smaller than that reached by the dijet data, the lower minimum
$p_T$ required for $W \rightarrow l \nu_l$ decays, allows these measurements to
probe lower values of $x$.  While the 5.02~TeV data for $W^\pm$ \cite{WpWm_5TeV}
and $Z^0$ production \cite{Z0_5TeV} were used in the EPPS16 global analysis,
the new $W^\pm$ measurements at 8.16~TeV \cite{WpWm_8TeV}
showed that EPPS16 gives a better
description of these data than older sets like nCTEQ that have not yet been
updated with the LHC data to guide them.

\section{Top Quarks}

\begin{figure}[htb]
\begin{center}
\includegraphics[width=0.45\textwidth]{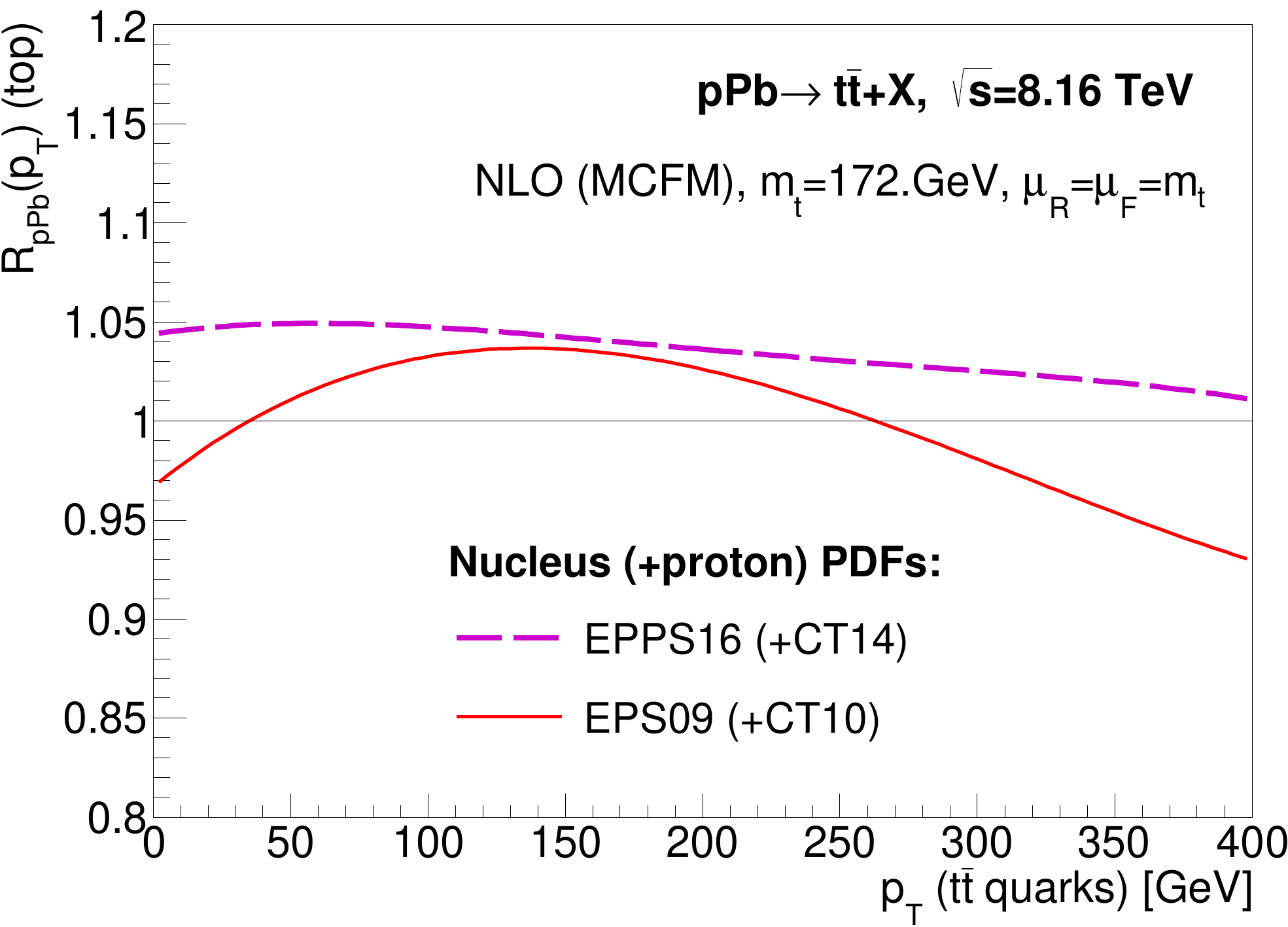}  
\includegraphics[width=0.45\textwidth]{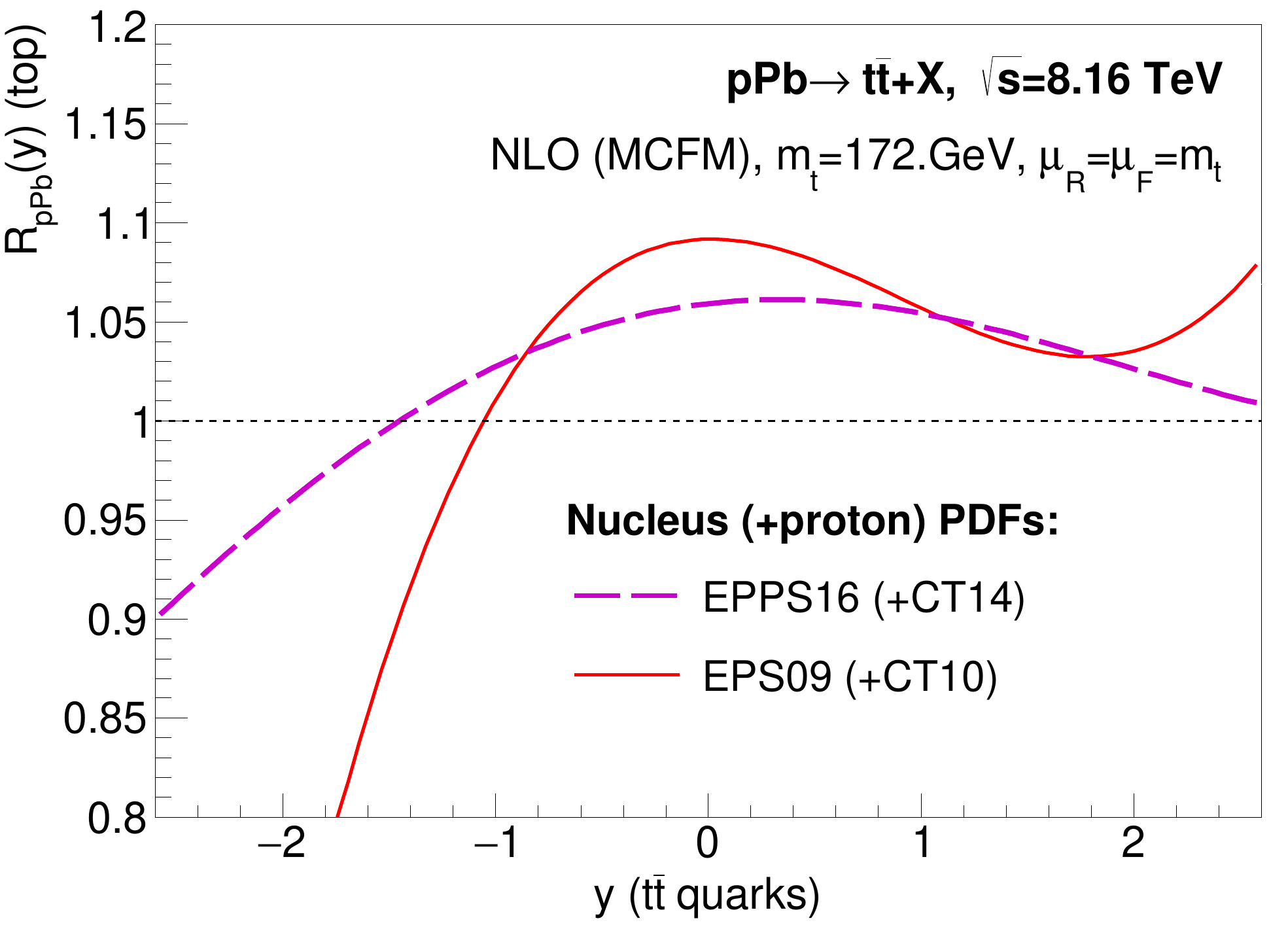}  
  \end{center}
\caption[]{Nuclear modification factor as a function of $p_T$
  (left) and rapidity (right) for $t \overline t$ production in the $\ell$+jets
  channel at $\sqrt{s_{_{NN}}} = 8.16$~TeV.
  Taken from Ref.~\protect\cite{Pred3}.
}
\label{fig:top}
\end{figure}

Top quark production in $p+$Pb collisions, the most massive, highest scale
probe so far for these collisions, was explored in a feasibility study by
d'Enterria {\it et al.} \cite{d'Enterria:2015mgr}
and also presented in Ref.~\cite{Pred3}.  The
measurement was carried out by CMS and reported in the lepton + jet channels:
$l +$jets, $\mu +$jets, and $e +$jets in Ref.~\cite{CMS_ttbar}.
So far, only a total
cross section in the available phase space can be reported.  Given the shorter
run time for $p+$Pb compared to $p+p$, the uncertainties on the data are
significantly larger for $p+$Pb relative to $p+p$.  The uncertainties on the
predicted cross sections are also larger since the nPDF uncertainties must also
be taken into account.  The additional sets in EPPS16 relative to EPS09 results
in the largest uncertainty band for this set.  Better constraints on the gluon
nPDF will help reduce these uncertainties in the future.  In addition, higher
statistics, either at the LHC or a future circular collider, will make it
possible to bin the data in different kinematic regions, making it possible to
compare to predictions such as those shown in Fig.~\ref{fig:top}.

\section{Summary}

With $p+$Pb collisions at the LHC, the study of perturbative probes of cold
nuclear matter in these collisions has entered a new era.  High statistics
studies of quarkonium and heavy flavors are available to probe the low $x$ and
moderate $Q^2$ region while high $p_T$ dijets offer the first clean probe of
nuclear gluon PDFs.  Gauge boson measurements are now mature enough to
distinguish between nPDF sets and separate nuclear effects on individual
parton densities than previously possible.  These measurements, along with the
first observations of top quark production in $p+$Pb collisions, show that
the parton distributions in nuclei are modified at every scale probed so far.

\section*{Acknowledgments}
I thank all my coauthors of Refs.~\cite{Pred1,Pred2,Pred3} for their
collegial collaboration over the years.  I also thank the organizers for their
kind invitation.
This work was performed under the auspices of the 
U.S. Department of Energy by Lawrence Livermore National Laboratory under 
Contract DE-AC52-07NA27344 and supported by the U.S. Department of Energy, 
Office of Science, Office of Nuclear Physics (Nuclear Theory) under contract 
number DE-SC-0004014.

\end{document}